\documentstyle[twocolumn,prl,aps,epsf]{revtex}
\begin{document}
\def \beq{\begin{equation}}
\def \eeq{\end{equation}}
\def \beqarr{\begin{eqnarray}}
\def \eeqarr{\end{eqnarray}}

\twocolumn[\hsize\textwidth\columnwidth\hsize\csname @twocolumnfalse\endcsname
\draft

\title{Ferromagnetic Transition in One-Dimensional Itinerant
Electron Systems
}
%Bonsinization and $\epsilon$ expansion}

\author{Kun Yang}
\address{
%National High Magnetic Field Laboratory and 
Department of Physics,
Florida State University, Tallahassee, Florida 32306
}

\date{\today}

\maketitle

\begin{abstract}

We use bosonization to derive 
the effective field theory that properly describes ferromagnetic 
transition in one-dimensional itinerant electron systems. The resultant theory
is shown to have dynamical exponent $z=2$ at tree level and 
upper critical dimension $d_c=2$. Thus one dimension is below the
upper critical dimension of the theory, and the critical 
behavior of the transition is controlled by an interacting fixed
point, which we study via $\epsilon$ expansion.
Comparisons will be made with the Hertz-Millis theory, which describes the 
ferromagnetic transition in higher dimensions.

\end{abstract}
]

Ferromagnetic transitions in itinerant electron systems are among the very 
first examples of quantum phase transitions studied theoretically\cite{hertz}.
In the approach pioneered by Hertz, one decouples the electron-electron 
interaction using Hubbard-Stratonovish transformation, integrates out the
fermionic degrees of freedom, 
and arrives at a Ginsburg-Landau-Wilson like free energy functional
that involves the ferromagnetic order parameter only, which are bosonic degrees
of freedom. The quantum nature of the theory lies in the (imaginary) time
dependence of the 
order parameter. This effective bosonic theory, known as the Hertz-Millis 
theory, was argued to have upper critical dimension $d_c=1$, thus the 
critical behavior of the transition is expected to be mean-field like for both
$d=2$ and $d=3$\cite{hertz,millis}.  
It has been pointed out recently, however, that the procedure of integrating
out gapless fermions may lead to subtle singularities in the expansion of the
resultant bosonic free energy functional in terms of the order parameter, or
its gradients\cite{bk}; such singularities may invalidate the power-counting
analysis of Hertz and Millis, and change the critical behavior of the 
transition\cite{bk}. The nature of the transition in two- and three-dimensional
itinerant electron systems is currently under extensive theoretical 
study\cite{bk}.

Comparatively speaking, 
much less attention has been devoted to the possibility of 
ferromagnetic transition in one-dimensional (1D)
electron systems, until recently. 
This is in part 
due to a theorem of Lieb and Mattis\cite{lieb}, which states that the ground
state is a singlet for certain classes of one-dimensional models with 
spin-independent interactions, thus rules out the possibility of ferromagnetism
in these models. The existence of ferromagnetic ground states in 1D models 
(that are
not dictated by the Lieb-Mattis theorem) was only established 
recently through numerical work\cite{daul}, in which it was also found that
the ferromagnetic transition is second-order, thus there is a quantum 
critical point. 
On the experimental side, the
so-called ``$0.7(2e^2)/h$" anomaly (or 0.7 anomaly) in the density-dependence
of the conductance of a 1D electron wire has attracted considerable attention;
one of the possible interpretations of this anomaly is
spontaneous 
magnetization of the electrons, which would give rise to plateau-like behavior
near $0.5(2e^2)/h$\cite{thomas}. This interpretation receives further 
support from the observation of similar behavior in the presence of magnetic
field near crossings of subbands with opposite spin polarization\cite{graham};
in this case the corresponding interpretation is spontaneous
pseudospin magnetization of the electrons, where the pseudospin index is 
actually the subband index. Inspired by these developments, a description of the
ferromagnetic phase of a 1D electron liquid is developed in 
Ref.\onlinecite{kopietz}, although the critical behavior
of the ferromagnetic transition was not studied in that work.

In this paper we study the
critical behavior
of ferromagnetic transition in 1D. We derive the effective bosonic field
theory that describes the transition using Abelian bosonization, and use 
renormalization group to study the critical behavior. The bosonization method,
which is specific to 1D systems and extremely powerful,
allows one to derive the effective bosonic theory in terms of the ferromagnetic
order parameter {\em without} having to integrate out gapless fermions; thus
the resultant theory does {\em not} suffer from singularities that the 
Hertz-Millis theory may encounter. A straightforward power-counting analysis
of the theory indicates that the dynamical exponent $z=2$ at the transition
(at tree-level), 
thus the upper critical dimension is $d_c=4-z=2$. As a result the theory is
{\em below} its upper critical dimension in 1D, thus the critical behavior
of the transition is controlled by an interacting fixed point. We study the
critical behavior using momentum shell renormalization group combined with 
$\epsilon$ expansion near $d_c=2$, at zero as well as low temperature.

{\bf Model and Bosonization} --- Consider the following Hubbard-like
Hamiltonian describing
interacting electrons in 1D:
\beq
H=\sum_{ij\sigma}t_{ij}c_{i\sigma}^\dagger c_{j\sigma}
+U\sum_{i}n_{i\uparrow}n_{i\downarrow}
+\sum_{ij\sigma\sigma'}V_{ij}^{\sigma\sigma'}n_{i\sigma}n_{j\sigma'},
\label{ham}
\eeq 
where $t_{ij}$ are the single electron hopping matrix elements, $U>0$ is an
onsite repulsion, 
and $V_{ij}^{\sigma\sigma'}$ represents  
(possibly spin-dependent) further neighbor repulsion; 
depending on the spin dependence 
of $V$ the system may possess either full Heisenberg (or O(3)) symmetry, or 
just Ising (or Z(2)$\times$U(1), the latter is responsible for conservation of 
$S_z^{tot}$) symmetry. 
It has been shown\cite{daul} that in the presence of further neighbor hopping
($t_{ij}$ with $|i-j|>1$), the Lieb-Mattis theorem no longer applies and a 
ferromagnetic ground state is stabilized for large enough $U$, when $V=0$.
One can also stabilize the ferromagnetic phase by having $V$ terms that are
spin-dependent and ferromagnetic; this possibility has been considered in the 
context of atoms trapped in 1D optical lattices\cite{pu}.
 
One of the most powerful methods of tackling
such 1D models is Abelian bosonization\cite{voit}. In this scheme, one 
takes advantage of the fact that in 1D,
all particle-hole excitations can be generated
by electron density and current
operators which satisfy bosonic commutation relations, 
and expresses both kinetic energy and interaction terms of $H$ in terms of 
the electron density and current operators. If one keeps terms that are 
quadratic and with least number of gradients
in the density and current operators, 
%which is justified (see below)
%in the weak coupling limit, 
one arrives at the familiar 
Luttinger liquid (LL) Hamiltonian, which describe decoupled spin and charge
excitations of the {\em paramagnetic phase}\cite{voit}:  
\beqarr
\label{hll}
&&H_{LL}=H_c+H_s;\\
&&H_c={1\over 2\pi}\int{dx}[\pi^2v_{Jc}\Pi^2_c(x)+v_{Nc}
(\partial_x\phi_c(x))^2];\\
\label{hs}
&&H_s={1\over 2\pi}\int{dx}[\pi^2v_{Js}\Pi^2_s(x)+v_{Ns}
(\partial_x\phi_s(x))^2].
\eeqarr
Here $\phi_c(x)$ and $\phi_s(x)$ are the charge and spin fields related to 
the charge and spin densities of the system:
\begin{equation}
\rho(x)={1\over \pi}\partial_x\phi_c(x),\hskip 0.5cm
S_z(x)={1\over 2\pi}\partial_x\phi_s(x);
\end{equation}
while $\Pi_\alpha$ are their conjugate fields satisfying
\begin{equation}
[\phi_\alpha(x), \Pi_{\alpha'}(x')]=i\delta_{\alpha\alpha'}\delta(x-x'),
\end{equation}
with $\alpha$ being $c$ or $s$. Physically $\Pi_\alpha(x)$ represents local
charge or spin current.  Clearly
the velocity parameters $v_{Nc}$ and $v_{Ns}$ parametrize the energy
cost of charge and spin density
fluctuations respectively, and are thus proportional to
the inverse charge and spin susceptibilities, 
while $v_{Jc}$ and $v_{Js}$ are proportional to the charge and spin
stiffness of the system respectively, as they measure the energy cost of
charge and spin current fluctuations.

We emphasize that $H_{LL}$
is an 
{\em approximation} of the original electron Hamiltonian Eq. (\ref{ham})
with generic single electron dispersion relation and two-body interaction.
For example, a nonlinear term of the form $\int{dx}\cos(\sqrt{8}\phi_s)$
that describes back scattering of electrons with opposite spins is neglected 
here. As pointed by Haldane\cite{haldane}, non-linearity in single-electron
dispersion gives rise to terms beyond quadratic order in $\Pi$ and $\phi$,
which represent interactions among the bosons.
Also the non-locality of electron-electron interaction (due to $V$ for example)
leads to
non-trivial wave-vector dependence in Fourier space, which
gives rise to terms that are
quadratic in $\phi$ but involve higher gradients. These terms, however, are
{\em irrelevant} in the renormalization group sense, {\em at the Luttinger liquid
fixed point} described by $H_{LL}$ (Eq. (\ref{hll})); they scale to zero in the long-wave length
and low-energy limit; their physical effect is to renormalize the parameters of
$H_{LL}$. Thus the long-wave length, low-energy properties of the system are
well described by $H_{LL}$, albeit with renormalized parameters. This is the
essence of the Luttinger liquid theory\cite{haldane,voit}.  
What we are going to see below however, is that some of the terms that are 
irrelevant and neglected at the Luttinger liquid fixed point
(which describes
the paramagnetic phase only) are crucial for 
the stability of the ferromagnetic phase, as well as
the ferromagnetic critical point;
they must be retained for a proper description of the ferromagnetic phase
as well as the transition. This should not be surprising, as operators
irrelevant
at one fixed may well be relevant at other fixed points\cite{yang}.

We now consider approaching the second order phase boundary from the 
paramagnetic side, by changing parameters in the Hamiltonian.
As we approach the critical point, the spin susceptibility $\chi$ diverges;
thus $v_{Ns}\propto 1/\chi\rightarrow 0$!
As we move further into the ferromagnetic phase, one expects $v_{Ns}$ to become
negative;
when this 
happens it becomes energetically favorable to have a non-zero expectation
value of $\partial_x\phi_s(x)$, which is the spontaneous magnetization. 
Physically this occurs because the gain of exchange energy from the 
magnetization overcomes the loss of kinetic energy,
as is standard in ferromagnetism in itinerant 
electron systems. Clearly with a negative coefficient for $(\partial_x\phi_s(x))^2$, $H_s$ is not stable, and higher order terms in gradients of $\phi_s(x)$
and powers of $\partial_x\phi_s(x)$ must be retained to maintain stability:
\beq
H'=\int{dx}[a(\partial^2_x\phi_s(x))^2+b(\partial_x\phi_s(x))^4+\cdots],
\eeq 
where in the weak coupling limit $a\approx (1/8\pi^2)\sum_j j^2(V^{+-}_{n,n+j}
-V^{++}_{n,n+j})$ and $b\approx {1\over 24\pi}
{d^3\epsilon(k)\over dk^3}|_{k=k_F}$ ($\epsilon(k)$ is the single electron 
dispersion).
%As discussed earlier, such terms are present in the bosonized Hamiltonian of 
%the system, but neglected on grounds that they are irrelevant at the Luttinger
%liquid fixed point. It is clear they must be retained for a proper 
%description of the ferromagnetic phase and transition\cite{yang}. 
It is clear that
the coefficient $a$ is positive for generic repulsive interactions 
when there is 
Ising anisotropy ($V^{+-} > V^{++}= V^{--}$).
On the 
other hand the sign of coefficient $b$ depends on the details of single 
electron dispersion; we assume $b$ to be positive here\cite{note}, 
so that the transition 
is second order which corresponds to the situation in Ref. \onlinecite{daul}.
Other possible terms that are allowed by symmetry can be shown to be irrelevant
at the ferromagnetic critical point to be studied below\cite{note1}.
Thus combining $H_s$ with $H'$, switching from Hamiltonian to action which is
more convenient for RG analysis, and after rescaling the spin field
($\phi=\sqrt{2a}\phi_s$), we arrive
at the following effective action:
\beq
S=\int{d\tau dx}\left[{1\over2}(\partial_\tau \phi)^2
+{1\over2}(\partial^2_x\phi)^2
+{r\over 2}(\partial_x\phi)^2
+u(\partial_x\phi)^4\right],
\label{action}
\eeq
where $\tau$ is imaginary time, $r=v_{Ns}/(4\pi a)$, and $u=b/(4a^2)$.
At mean field level, 
the ferromagnetic transition occurs at $r=0$. 
We note that the last three terms in Eq. (\ref{action}) (or the static parts)
take the usual form of Landau theory, upon identifying $\partial_x\phi$ as the
local magnetization $m(x)$: ${r\over 2}m^2+{1\over2}(\partial_xm)^2+um^4$.
The first term in Eq. (\ref{action}) controls the fluctuations along the 
imaginary time direction, and is responsible for the quantum nature of the
theory.
The effective action (\ref{action}) is the basis of our study of the critical behavior of the ferromagnetic 
transition, which we now turn to.

{\bf Renormalization Group and Critical Behavior at $T=0$} ---
We perform a renormalization group (RG) analysis of the 
action (\ref{action}), and as is standard in such analysis, we treat the spatial
dimension $d$ as a continuous variable, even though the action 
(\ref{action}) is 
derived in 1D. We perform the following space-time transformation with 
scale factor $s>1$:
$x'=x/s, t'=t/s^z$, and $\phi'(x')=\phi(x)s^\Delta$, that leaves the first two
terms in (\ref{action}) invariant. This leads to $z=2$ and $\Delta=d/2-1$,
which in turn lead to the (tree-level) scaling relation for $r$ and $u$:
$r'=rs^2$ and $u'=us^{2-d}$. We thus find $u$ is {\em relevant} below the 
upper critical dimension $d_c=2$, and the critical property of the transition
is controlled by an interacting fixed point for $d < 2$. Obviously this fixed 
point is very different from the Luttinger liquid fixed point (which is 
non-interacting and has dynamical exponent $z=1$), even though we are in 1D. 
In the following
we study the critical property using momentum shell RG combined with 
$\epsilon=2-d$ expansion, and hope it gives a reasonable description at $d=1$. 

We assume there is a momentum cutoff $\Lambda$, while no such cutoff exists in
frequency. Integrating out modes with $\Lambda/s < k < \Lambda$ and arbitrary
frequency at one loop level, 
we obtain the following flow equations for $r$ and $u$ for small $\epsilon$:
\beqarr
{dr\over d\log s}&=&2r+{3u\over \pi}(\Lambda^2-{1\over 2}r);\\
{du\over d\log s}&=&\epsilon u-{9\over 2\pi}u^2.
\eeqarr
From these we obtain the fixed point (to order $O(\epsilon)$):
$u^*=(2\pi/9)\epsilon$ and 
$r^*=-(\epsilon/3)\Lambda^2$. This interacting fixed point
controls the critical property of the 
transition, at which the RG dimension of the (relevant) tuning parameter $r$ is
$y_r=2(1-\epsilon/6)$. From $y_r$ as well as the fact that the field $\phi$ 
receives no anomalous dimension ($\eta=0$) to order $O(\epsilon)$, we 
determine the critical exponents: correlation length exponent $\nu=1/y_r\approx
(1/2)(1+\epsilon/6)$; susceptibility exponent $\gamma=(2-\eta)\nu\approx
1+\epsilon/6$; magnetization exponent $\beta=\nu d/2\approx 1/2-\epsilon/6$;
and field exponent $\delta\approx 3+\epsilon$. 
%The procedures of obtaining these exponents from $y_r$ and $\eta$ are very 
%similar to those of the $\epsilon$ expansion for the $\phi^4$ theory near 
%four-dimensions, and amazingly, the exponents obtained here are the same as 
%those of the Wilson-Fisher fixed point to order $O(\epsilon)$\cite{wilson}. 
To order $O(\epsilon^2)$ the field $\phi$ receives a non-zero anomalous 
dimension $\eta$, which also leads to a correction to the dynamical exponent
$z=2-\eta/2$\cite{sachdev}. 

{\bf RG and Critical Behavior at Finite $T$} ---
The phase transition occurs at $T=0$ only. However the quantum critical point
has very significant influence on the thermodynamic properties at finite $T$,
if the system is sufficiently close to the critical point. 
%Since all experiments are performed at finite $T$, the critical behavior at low
%but finite $T$ is actually more important.
More specifically,
as we will see below, finite temperature introduces a thermal length scale 
$\xi_T\sim T^{-1/z}$, and depending on its interplay with the correlation length
$\xi$ of the system at $T=0$,
one can divide the temperature-coupling space into three regions,
which are separated from each other by two crossover lines of the form
$T\sim |r-r^*|^{z\nu}$\cite{chn}:

(i) Quantum Disordered: $r > r^*$ (where $r^*$ is the critical coupling) and    
$\xi_T > \xi$, in which the system behaves as an ordinary Luttinger liquid;
for example the specific heat $C\sim T$ and susceptibility $\chi\sim constant$.

(ii) Renormalized Classical: $r < r^*$ and $\xi_T > \xi$, the system behaves
like a ferromagnetic Luttinger liquid\cite{kopietz} whose magnetic order
is suppressed by thermal fluctuations; here depending on the symmetry there can 
be two types of behavior: (a) if the system possesses full Heisenberg symmetry,
the gapless transverse spin fluctuation with spectrum $\omega\sim k^2$ gives 
rise to specific heat $C\sim\sqrt{T}$, while the susceptibility
$\chi\sim 1/T^2$\cite{fisher}; (b) if the system possesses Ising symmetry only,
then the transverse spin fluctuation is gapped, the longitudinal fluctuation 
with linear spectrum (the usual Luttinger liquid behavior) gives 
rise to specific heat $C\sim T$, while the susceptibility
$\chi\sim\exp(J/T)$, which is the usual behavior of an Ising ferromagnet
($J$ is an energy scale of order the domain wall energy of the Ising ferromagnet).

(iii) Quantum Critical: $\xi_T < \xi$, in which the thermodynamic property is
controlled by the quantum critical fixed point studied above. In the following we focus on the quantum critical region. To study finite $T$ properties, we
need to generalize the RG flow equation in the presence of finite temperature.
The temperature $T$ determines the range of imaginary time: $0 < \tau < 1/T$.
Thus under scaling $T$ scales as: $T'=Ts^z$. The flow equation of $r$ is modified to be
\beq
{dr\over d\log s}=2r+{3u\over \pi}(\Lambda^2-{1\over 2}r)
\coth{\Lambda^2\over 2T},
\label{tflow}
\eeq
which leads to a flow {\em away} from the fixed point value $r=r^*$ due to 
finite $T$ (flow of $u$ is negligible if initially $u=u^*$). 
Integrating Eq. (\ref{tflow}) till
$r-r^*\sim \Lambda^2$, at which scale the system is far away from criticality
and the correlation length
$\xi(s)\sim 1/\Lambda$, we find that the temperature $T$ sets a correlation
length 
\beq
\xi_T=s\xi(s)\approx {1\over\Lambda\epsilon^{(1/y_r)}}\left({T\over\Lambda^2}\right)^{-1/z}\propto
T^{-1/z}
\eeq
as anticipated.
This leads to the temperature dependence of susceptibility 
$\chi\sim\xi^{2}\sim1/(\epsilon T)$ in the quantum critical region, to order
$O(\epsilon)$. To determine the behavior of specific heat, we need the singular
contribution to the free energy from critical fluctuations, which obeys the
hyper-scaling law for theories below their upper critical dimensions:
\beq
F\sim T^{1+d/z}\Phi(|r-r^*|^{z\nu}/T),
\label{free}
\eeq
where $\Phi(x)$ is a universal scaling function. From Eq. (\ref{free}) we
immediately obtain $C\sim T^{1-\epsilon/2}$ in the quantum critical region.  

{\bf Summary and Discussion} --- In this work we developed a bosonic field 
theory that describes ferromagnetic transition in 1D itinerant electron systems,
based on Abelian bosonization. This approach is quite different from that of
the Hertz-Millis theory, because in principle the bosonization procedure keeps
all the degrees of freedom of the original fermionic system, and allows us to 
arrive at a bosonic theory {\em without} integrating out gapless fermionic 
degrees of freedom, as was done in the Hertz-Millis theory. We thus believe 
the theory developed here is free of the possible singularities associated 
with integrating out gapless fermions.

The bosonic theory developed here, when generalized to arbitrary dimensions, was
found to have upper critical dimension $d_c=2$. Thus for $d=1$, which is where
the theory applies, the system is below its upper critical dimension, and the
universal 
critical behavior of the transition is controlled by an interacting fixed 
point, which we have studied in some detail using $\epsilon$ expansion. This is
again quite different from the Hertz-Millis theory, which is above its 
upper critical dimension $d_c=1$ for $d=2$ and $d=3$ where it applies; there
the critical behavior is controlled by a Gaussian fixed point, and some of the
critical properties are non-universal due to the presence of dangerously 
irrelevant operators. 

The Abelian bosonization procedure can be applied to systems either with full
Heisenberg symmetry or Ising symmetry only. However since it does not exhibit
the Heisenberg symmetry explicitly, this symmetry is most likely lost due to the
approximate nature of the derivation and treatment of the bosonic theory.
Thus the critical behavior discussed above probably applies to systems with
Ising symmetry only. In solid state systems the Heisenberg symmetry of electron
spins are often reduced to Ising due to the ubiquitous spin-orbit coupling;   
in systems with pseudospin transitions the Heisenberg symmetry is absent in
the first place. Thus the results presented here are highly relevant. 
Nevertheless it would be highly desirable to maintain the Heisenberg symmetry
when present, using for example non-Abelian bosonization, and study if and how
the extra symmetry affects critical behavior of the transition. We leave this
for future investigation.    

In an earlier work, Sachdev and Senthil\cite{ss} studied ferromagnetic 
transitions in lattice rotor models, and arrived at an effective action 
very similar
to Eq. (\ref{action}); the only difference being the field is complex in their
work, which is crucial for the Heisenberg (O(3))
symmetry of their model. If the
symmetry were reduced to Z(2)$\times$U(1), the corresponding theory would  
involve a real field, then the theory becomes
identical to Eq. (\ref{action})\cite{sachdev}. 
It was conjectured\cite{sachdev,ss} that this action properly describes 
ferromagnetic transition in 1D itinerant electron systems. It is remarkable that
two very different approaches lead to the same effective theory for the 
transition. 

Incidentally, a 2D version of the action (\ref{action}) was used
to study transitions between different valence bond solid states
in 2D recently\cite{ashvin}. The physics of these transitions are very different
from the one discussed here however.

This work was initiated while the author was visiting
the Max Planck Institute for Physics
of Complex Systems in Dresden, during her Workshop on
Quantum Phase Transitions.
He benefited from stimulating discussions with
Peter Kopietz. He is particularly grateful to Subir Sachdev and T. Senthil for
explaining to him the connection between Ref. \onlinecite{ss} and the present
work, and bringing Ref. \onlinecite{ashvin} to his attention.
This work was supported by NSF grant No. DMR-0225698.


\begin{references}
\bibitem{note} If $b$ is negative then
higher order terms in $\partial_x\phi_s(x)$ must be retained, which will lead to
a first-order ferromagnetic transition. Only even powers of $\partial_x\phi_s(x)$ are present as guaranteed by spin symmetry.
\bibitem{note1} In fact there are also non-linear terms of the form
$\int{dx}(\partial_x\phi_c(x))(\partial_x\phi_s(x))^2$ etc 
that actually couple spin and charge sectors; since the transition happens in 
the spin sector, the charge sector
can be integrated out which generates terms that are already present in
$H_s$ and $H'$. 
No singularities are generated here.
\end{references}
\end{document}